\documentclass[useAMS,usenatbib]{mn2e}

\title[Ion-proton pulsars]{
Antenna source of radio-frequency emission in ion-proton pulsars}
\author[P. B. Jones]{P. B. Jones\thanks{E-mail:
p.jones1@physics.ox.ac.uk}\\
University of Oxford, Department of Physics, Denys Wilkinson Building, 
Keble Road, Oxford OX1 3RH, England\\}
\begin{document}

\date{}

\pagerange{\pageref{}--\pageref{}} \pubyear{}

\maketitle

\label{firstpage}

\begin{abstract}
The growth of a longitudinal or quasi-longitudinal Langmuir mode in the outward-moving beam of ions and protons that forms the open sector of an ion-proton pulsar magnetosphere radiates as an analogue of an end-fed high-impedance horizontal straight-wire antenna an integral number of half-waves in length. The radiation has, broadly, the energy flux, linear polarization and spectral index that are widely observed: also, the notch phenomenon seen in some integrated pulse profiles occurs naturally. The new field of pulsar observations below 100 MHz may lead to productive tests of the radio emission mechanism.

\end{abstract}


\section{Introduction}

The physics of pulsar radio emission has remained an unsolved problem for more than half a century since the first observation by Hewish et al (1968).  It is our belief that the principal reason for this has been the universal assumption of an electron-positron plasma formed above the polar cap.  An extensive critique of this work in the previous millennium has been published by Melrose \& Gedalin (1999). Evidence for pair creation has been conspicuous by its absence: in long-period pulsars such as J2144-3933 (Young, Manchester \& Johnston 1999) it appears impossible assuming the known cross-sections of quantum electrodynamics. More generally, the side-effects of high-multiplicity pair production have not been seen.
Pair production in millisecond pulsars (MSP) must be accompanied by $\gamma$-ray production coincident in observed longitude with radio emission (see Jones 2021); also areas of neutron-star surface with raised thermal temperatures characteristic of reverse-flow electrons or positrons have not been observed in the same coincidence (see, for example, Rigoselli et al 2022).  The author is unaware of any counter-examples.

An ion-proton model has been described in previous papers (Jones 2016, 2020a) for neutron stars  having positive polar-cap corotational charge densities.  In brief, ions accelerated from the thin atmosphere at the polar cap surface undergo photo-electric transitions  in the whole-surface blackbody radiation field of the neutron star.  The reverse-accelerated electrons screen the acceleration field $E_{\parallel}$ and, in forming electromagnetic showers, create protons through the decay of the nuclear giant-dipole states of photo-nuclear reactions.  Protons diffuse to the neutron-star surface and, having the higher charge-to-mass ratio, are accelerated in preference to ions.  The system is self-regulating to a high degree, $E_{\parallel}$ being screened by reverse electrons so that ions and protons have only small to modest Lorentz factors appropriate for the required photo-electric transition rates.  It is not necessarily stable in the sense of being time-independent.
At an instant of time when the proton flux in a given locality falls below that consistent with the existing $E_{\parallel}$, ions are again accelerated and the process is repeated. Thus a polar cap at any instant generally has both areas of ion plus proton emission and areas of pure proton emission.  Outward acceleration of the ion-proton plasma beam is subject to the growth of a longitudinal Langmuir mode reaching large amplitudes at altitudes $z$ above the polar cap less than the neutron-star radius $R$.

Considerable energy is in principle available because two particle types of different charge-to-mass ratio accelerated as a beam through a common potential difference are not in their lowest energy state at a fixed total momentum, which would be one in which both particles have a common velocity.  Our previous assumption (Jones 2017) has been that a movement towards the lowest energy state would be by coupling with the radiation field through the development of non-linearity and possibly turbulence in the Langmuir mode.
But the nature of turbulence is constrained because the current fluctuations of the mode are always parallel with the local magnetic flux ${\bf B}$ and calculation of the vector potential ${\bf A}$, in the Lorentz gauge, depends only on the physical current and not on scalar potential fluctuations that occur through the continuity condition.

The emission region can be regarded as a black-box of which the following properties are required.

(i)	  The Langmuir-mode frequency is low immediately above the polar-cap, in many cases less than 100 MHz, so the emission process must include the development of smaller-amplitude higher wave-numbers.

(ii)  The direction and linear polarization of the electric vector must broadly conform with the historic observations of Radhakrishnan \& Cooke (1969).

(iii) Widths of integrated pulse profiles are roughly frequency-independent except at very low frequencies indicating an emission process confined to a modest interval of altitude.  This is also necessary if the millisecond pulsars  are assumed to emit by the same process.

(iv)  Some degree of coherence is needed to account for the observed luminosities: also for the fine detail observed in some high-quality integrated profiles.  Here coherence means a definite phase relation between the observer and each point in the emitting region.  Its effect is determined by the observer angle $\theta_{0}$ and the nature and structure of the current. It is likely that such coherence can be associated only with a mode, the Langmuir mode in the present case.

(v)	  The problem of a beam is three-dimensional and there must be large transverse gradients in many parameters. The cross-sectional area of the beam could conceivably sub-divide into a number of different areas lacking the relative coherence described in (iv). this would be consistent with the sub-pulse formation which is observed and indicates that coherence in directions parallel with ${\bf B}$ is more important than in the cross-sectional plane.

The functioning of the emission process described here is in many respects unusual. Therefore, it seems appropriate to give a brief introduction to the contents of Section 3.

\section{The Langmuir mode}

  The source of the radiation is here taken to be the current density distribution of a Langmuir mode confined to the flux lines of the open magnetosphere above the polar cap.  This set of flux lines is neither homogeneous nor one-dimensional and these facts are essential to the arguments given in Section 3. The magnetosphere above the polar cap consists of the open sector, defined as the narrow column of magnetic flux lines crossing the light cylinder, surrounded by the closed sector.  They are distinct plasmas separated, it is usually presumed, by a well-defined boundary condition.  The open sector contains those processes described in Section 1 in an outward moving two-component plasma of ions and protons: the closed sector is usually assumed to be charge-separated, of a single ion component, corotating and with no outward velocity.
  
 This paper assumes that, provided its wavelength is an order of magnitude smaller than the column radius $u_{m}$, an unstable Langmuir mode forms in the open sector with growth rate approximately as in the homogeneous one-dimensional plasma case.  The dispersion relation is found by solution of the zero-temperature Maxwell-Vlasov equations for a relativistic multi-component plasma in this latter case.  It is, for the longitudinal mode,
\begin{eqnarray}
D_{zz}(\omega) = 1 + \sum_{i}\frac{m_{i}\omega^{2}_{i}}{q}\int^{\infty}_{-\infty} dp
\frac{\partial f_{i}}{\partial p}\frac{1}{\omega - qv_{i}(p) +i\epsilon} = 0
\end{eqnarray}
with the imposition of this condition on the dielectric tensor.  The functions $f_{i}$ are the particle distribution functions, normalized to unity. In the ion-proton model, this is $f_{i} = \delta(p - p_{i})$ which is not an approximation. For $i = 1,2$ the relativistic dispersion relation becomes
\begin{eqnarray}
D_{zz} = 1 - \frac{\omega^{2}_{s1}}{(\omega - qv_{1})^{2}} - \frac{\omega^{2}_{s2}}
{(\omega - qv_{2})^{2}} = 0
\end{eqnarray}
in which $\omega^{2}_{i} = 4\pi n_{i}Z^{2}_{i}{\bf e}^{2}/m_{i}$, is the plasma frequency for particle (i) and  $\omega^{2}_{si} = \omega^{2}_{i}/\gamma_{i}^{3}$. The velocities and Lorentz factors are $v_{1,2}$ and $\gamma_{1,2}$; the particle masses $m_{1,2}$ are $A_{1,2}m_{p}$ in terms of the proton mass and the nuclear mass number $A$.  The angular frequency $\omega$ and all other parameters are measured in the observer frame, which is assumed here to be identical with the corotating frame.
At least two components of different charge-to-mass ratio are essential for the existence of the mode.
Their number densities $n_{1,2}$ depend on the ion-proton ratio at any instant but give a charge density $(n_{1}Z_{1} + n_{2}Z_{2}){\rm e} = N{\rm e}$ which is close to the Goldreich-Julian density, but modified by the presence of the small reverse flux of photoelectrons.

The simplicity of the ion-proton model has, as its basis, the reduction of the familiar equation (1) to equation (2). This is not an approximation because particles of distinct charge-to-mass ratio, accelerated from rest through a common potential difference, have distinct velocities and if baryonic, have negligible energy losses from bremsstrahlung. Landau damping, represented by the imaginary part of the integral  in equation (1), is not present.

The existence of stray electron-positron pairs in the open sector does not require the modification of equation (2). The acceleration field $E_{\parallel}$ is of the order needed to produce screening photoelectrons: proton Lorentz factors are moderately relativistic, of the order of $\gamma_{p} \approx 10$.  This would produce electrons and positrons with Lorentz factors of the order of $10^{4}$ whose contribution to equation (2) would be negligible.

The quartic equation (2) can have two real solutions and a conjugate pair, the necessary condition being,
\begin{eqnarray}
D_{zz}({\tilde{\omega}}) = 1 -\frac{C}{q^{2}} < 0
\end{eqnarray}
with,
\begin{eqnarray}
{\tilde{\omega}} = q\left(\frac{v_{1}\omega_{s2}^{2/3} + v_{2}\omega_{s1}^{2/3}}
{\omega_{s1}^{2/3} + \omega_{s2}^{2/3}}\right),
\end{eqnarray}
and
\begin{eqnarray}
C =  \frac{1}{(v_{2} - v_{1})^{2}}\left(\omega_{s1}^{2/3} + \omega_{s2}^{2/3}\right)^{3}.
\end{eqnarray}
The growth rate of the mode has been investigated previously (Jones, 2012a,b) and applied to a model for the long-period pulsar J2144-3933 (Jones 2022).  A continuous band of solutions $\omega(q)$ satisfying equations (3) - (5) exists but with a maximum growth rate at a specific $\omega$.  Consequently, in a homogeneous plasma,  there is a spectrum of $\omega$ dependent on the manner of initial seeding of the mode, which evolves on propagation, becoming narrower and centred on the specific frequency of greatest growth rate.  It is realistic to define this as the mode frequency.  The inhomogeneous case can be represented by the notional division of the plasma into layers of thin parallel homogeneous sheets and considering the thin limit, as in the present problem, the decrease of the Goldreich-Julian density (Goldreich \& Julian 1969) with increasing altitude above the polar cap.  At any interface, the components of the electric field parallel with its plane must be continuous functions at the interface at all times. Thus the frequencies in the spectrum of the electric field are necessarily unchanged. But the amplitudes of individual frequencies in the normal field component are not so constrained and the field as a whole can change as a consequence of the incremental change in dispersion relation parameters.  Consequently, the shape of the frequency spectrum changes so that the maximum growth-rate frequency, defined here as the mode frequency, is reduced as the Goldreich-Julian density decreases with altitude above the polar cap. This is in contrast with stable plasma modes of discrete frequency for which the above argument has no applicability, and for which $\omega$ remains constant.

Quasi-longitudinal unstable Langmuir modes also exist (see Asseo, Pelletier \& Sol, 1990) and couple directly with the radiation field, but their growth rates are lower than for the equivalent longitudinal case.  The wave-vector ${\bf q}$ is not a constant of motion and it is anticipated that the mode would converge to the longitudinal case.

The longitudinal mode restricted to the open-sector plasma radiates, unlike the mode in a homogeneous one-dimensional plasma, and the radiation is restricted to directions within a small-angle cone of axis parallel with ${\bf B}$.  Almost invariably, the cone lies within the open sector of the magnetosphere.

 In the following Section, computations of the distant magnetic flux density from the Langmuir-mode current density is as in free space.  The justification is that O-mode refractive indices are extremely close to unity in baryonic plasmas: that derived from the dielectric tensor given by Beskin \& Philippov (2012) and valid for relativistic beams is
\begin{eqnarray}
n_{O} = 1 - \frac{\omega_{p}^{2}\sin^{2}\theta}{2\gamma^{3}\omega^{2}(1 - \cos\theta +1/2\gamma^{2})^{2}}
\end{eqnarray}
at an angle $\theta$ with ${\bf B}$.  Here $\omega_{p}$ and $\gamma$ represent the plasma frequency and Lorentz factor of the particle beam, evaluated in the observer frame. For fixed $\gamma$ and $\omega$, it is seen that $|n_{0} - 1| \propto m^{-1}_{i}$ .  Our assumption is that the condition $|n_{0} - 1| \ll 1$ can be replaced adequately by $n_{0} = 1$. The problem then is simply that of obtaining distant solutions of equation (7).  Melrose, Rafat \& Mastrano (2021) also arrived at the above condition (see Section 5.4 of their paper) when considering the propagation of O-modes in the pulsar magnetosphere for a new alternative emission mechanism based on rotation-driven superluminal longitudinal plasma waves.

Finally, with the E-mode $n_{E} = 1$, there is sufficient birefringence to account for the circular polarization observed in a great many pulsars (see Jones 2020b). Hence, in Section 3, we assume that O-modes propagate as $\omega = ck$.

\section{The luminosity and spectrum}

Some simplifications are necessary and so we assume that the beam described in Section 1 is circular in cross-section and its centre is coincident with the z-axis, the magnetic axis of a dipole field. It is also assumed that the emission of radiation starts at a definite altitude $z_{0} = R(\eta_{0} - 1)$ above the neutron star surface and ceases at $z_{f} = R(\eta_{f} - 1)$, where $R$ is the neutron star radius and $\eta$ is the radius in units of $R$.          
At radiation field distances we use spherical polar coordinates ($r,\theta,\phi$)with origin at the centre of the neutron star: the observer direction is in the x-z plane at an angle $\theta_{0}$ with the z-axis.  The beam is of uniform constitution in a cross-section perpendicular to the z-axis but its radius is $u_{m}(\eta)\propto\eta^{3/2}$ for a dipole field. The cross section is described in circular polar coordinates ($u,\phi$).  In the present case, the current distribution produces a vector potential component $A_{\theta}$ in the radiation field,
\begin{eqnarray}
A_{\theta}({\bf r},t) = \frac{1}{c}\int^{z_{f}}_{z_{0}}d{\bf r}^{\prime}(J_{x}\cos\theta_{0} - J_{z}\sin\theta_{0})
\frac{\exp(ik|{\bf r} - {\bf r}^{\prime}|)}{|{\bf r} - {\bf r}^{\prime}|},
\end{eqnarray}
(Jackson 1962) in which
 $|{\bf r} - {\bf r}^{\prime}| = r - \hat{\bf n}\cdot{\bf r}^{\prime}$ with
 $\hat{\bf n} = (\sin\theta_{0}, 0, \cos\theta_{0})$ being a unit vector parallel with the direction of observation and ${\bf r}^{\prime} = (u\cos\phi, u\sin\phi, z)$. Neither cut-off in $z$ is present in the real problem considered here but they are adopted for convenience.  Growth of the longitudinal Langmuir mode produces a current density component ${\bf J}$ which is resolved into components $J\sin\chi\cos\phi, J\sin\chi\sin\phi, J\cos\chi$ with phase $\exp(i(qz + qu\sin\chi/2 - \omega t))$ and $J = N(z)e\delta v$, where $Ne$ is the charge density, $\propto \eta^{-3}$. Here $\delta v$ is the mode velocity fluctuation  of wave number $q$ and angular frequency $\omega$. The frequency is $\propto N^{3/2}$. 
 
The magnetic flux density in the radiation field is then,
\begin{eqnarray}
B_{\phi}({\bf r},t) = \frac{\pi u_{m}^{2}J}{rc}\int^{z_{f}}_{z_{0}} dz(ik)\exp(i(kr - \omega t))
\\   \nonumber
\exp(iqz(1 - \cos\theta_{0}))F(\chi, \theta_{0}),
\end{eqnarray}
with,
\begin{eqnarray}
F = \frac{1}{\pi u_{m}^{2}}\int^{u_{m}}_{0}u du \int^{2\pi}_{0} d\phi  \\  \nonumber
(\chi\cos\phi - \sin\theta_{0}) \exp(iqu(\sin\chi/2 - \sin\theta_{0}\cos\phi)),
\end{eqnarray}
following the arguments of Section 2. In our initial evaluation of equations (8) and (9) we shall assume $q = k$.  (As a special case we can set $\chi = 0 $  and assume that $u_{m}$ is constant and  small so that the wavenumber $k$ is also constant and that $F = -\sin\theta_{0}$.  
Completing the integral over $z$ we find that the radiation-field magnetic flux density is,
\begin{eqnarray}
B_{\phi} = \frac{ -ik\pi u_{m}^{2}J_{z}}{rc}\exp(i(kr - \omega t))\sin\theta_{0}
    \nonumber   \\    
\left(\frac{\exp(ik(z_{f} - z_{0})(1 - \cos\theta_{0}))- 1)}{ik(1 - \cos\theta_{0})}\right)
\end{eqnarray}
This has maxima for $k(z_{f} - z_{0})(1 - \cos\theta_{0}) = n\pi$ for odd integers $n$,
and zeros for even integers.  The terrestrial counterpart of this case is well known: an end-fed high-impedance horizontal straight-wire antenna, an integral number of half-waves in length. The maxima in the bracketed quantity in equation (10) are
$2i(z_{f} - z_{0})/n\pi$ and so the strongest lobes in the radiation pattern are close to being parallel with the wire.)

But in the present problem, the angular frequency of the mode is a function of particle number density, $\propto N^{1/2}$.  Also $z$ is not a quantity to which observables relate. Thus given our assumption of adiabatic change in $k$, we change variable from $z$ to $k$,
\begin{eqnarray}
z = R(\eta - \eta_{0}) = R\eta_{0}(k_{0}^{2/3}k^{-2/3} - 1).
\end{eqnarray}
We also assume that after emission and for a given $\omega$, the radiation wavenumber $k$ changes adiabatically to its free-space value. The products $ku_{m}$ and $J u_{m}^{2}$ are independent of $k$ and can be replaced by their values at $\eta_{0}$. We assume constant $\theta_{0}$ for which $F(\chi,\theta_{0})$ is simply a function of $\chi$.  The radiation field $B_{\phi}$ is then,
\begin{eqnarray}
B_{\phi}(r - ct) = C \int^{k_{f}}_{k_{0}} dk H(k) \exp(ik(r - ct))
\end{eqnarray}
in which,
\begin{eqnarray}
C = -\frac{2iR\eta_{0}}{3rc}\pi J(\eta_{0}) u_{m}^{2}(\eta_{0})
\end{eqnarray}
and $H = FG$ with,
\begin{eqnarray}
G = k_{0}^{2/3}k^{-2/3} \nonumber  \\
\exp(iqR\eta_{0}(1 - \cos\theta)(k_{0}^{2/3}k^{-2/3} - 1)).
\end{eqnarray}
The Fourier transform of $B_{\phi}(r - ct)$ is $B_{\phi}(k) = 2\pi CH(k)$. Because the radiation field electric vector is ${\bf E}_{\theta} = {\hat{\bf n}}\times{\bf B}_{\phi}$,  from Parseval's theorem, and integrating over unit time, we find the energy flux per unit area and time at distance $r$ is,
\begin{eqnarray}
\frac{c}{8\pi}\int dt |B_{\phi}(r - ct)|^{2} = \frac{c}{8\pi} \left\langle|B_{\phi}(r - ct)|^{2}\right\rangle     \nonumber   \\
		= -\frac{1}{8\pi} \int^{k_{f}}_{k{0}}dk |B_{\phi}(k)|^{2}.
\end{eqnarray}
A trial evaluation of equation (15) can be made by assuming the black-box to be a void, that is, by assuming the current density $J$ to be that of the linear Langmuir mode. For convenience, we evaluate $B_{\phi}$ at observer point $(r - ct) = 0$. The initial and final altitudes are $\eta_{0} =1.5$ and $\eta_{f} = 2.5$. One half of the beam (Goldreich-Julian) charge density consists of protons and we assume that they have Lorentz factor $\gamma_{p} = 9$. On this basis, the ion component of the beam, having the same charge density and chosen to consist of ions with charge Z and mass number  $A = 3Z$, has Lorentz factor $\gamma_{A,Z} =3.7$.  At the lower altitude $\eta_{0}$, evaluation of the constant $C$ gives a mode frequency upper limit,
\begin{eqnarray}
\nu_{0max} < 110 \left(\frac{B_{12}}{P}\right)^{1/2}
\hspace{2mm} {\rm MHz}.
\end{eqnarray}
Hence we shall adopt a frequency suitably below this for evaluation, thus replacing $110$ by $60$ MHz.
The $B_{12}$ factor here is the polar-cap surface field in units of $10^{12}$ G. The period is taken as $P = 1$ s and the neutron-star radius $R = 1.2 \times 10^{6}$ cm.  The angular distribution is determined by the function $G$ and to some extent $F$. We estimate a full width at half maximum of $2.8$ degrees and an energy flux of,
\begin{eqnarray}
3\times 10^{32}\left(\frac{1}{r^{2}}\right)\left(\frac{B_{12}}{P^{2}}\right)^{2}\left|\frac{\delta v}{c}\right|^{2} \hspace{3mm} {\rm erg}\hspace{2mm} {\rm cm}^{-2}{\rm s}^{-1},
\end{eqnarray}
when evaluated at $\theta_{0} = 4\times 10^{-2}$ radians.  The luminosity integrated within a cone of solid angle $5\times 10^{-3}$ steradians centred on the magnetic axis is approximately $1.5\times 10^{26}$ ergs s$^{-1}$ for $B_{12} = 1$, $P = 1$ s and $|\delta v/c| = 10^{-2}$.  At a distance of $1$ kpc, the signal strength is approximately $3\times 10^{3}$ mJy and the radio-frequency energy created per unit positive charge in the beam is $60$ MeV.

The factor $\delta v$ is a linear combination of the two fluctuations $\delta v_{p}$ and $\delta v_{A,Z}$ in protons and ions. They are related to fluctuations in Lorentz factor by $\delta v_{p} = \delta\gamma_{p}/(v_{p}\gamma_{p}^{3})$ etc and in the relativistic limit their ratio is $A^{2}/Z^{2}$.  Thus radio-frequency production requires moderately relativistic ions and protons, as in the ion-proton model, of which the ion fluctuation is likely to be the larger contributor to $\delta v$.

This is an interesting order of magnitude of luminosity and is approximately proportional to the spin-down energy loss-rate.  Also the radiation field $E_{\theta}(r - ct)$ is entirely consistent with the polarization seen by Radhakrishnan \& Cooke (1969), and the spectral index of approximately $-4/3$ is broadly consistent with that observed. 
The angular distribution is more narrow than that observed (see Posselt et al 2021) but would be broadened if the coherence length interval of the z-axis, $R(\eta_{f} - \eta_{0})$ were reduced.  The fact that $B_{\phi} = 0$ at $\theta_{0} = 0$ is a consequence of our cylindrically-symmetric assumed structure.  Hints of it appear in some profiles where two very sharp and closely-spaced peaks are seen. Generally, these are referred to as notches and occur in some normal pulsars and in some MSP (see McLaughlin \& Rankin 2004; also Dyks, Rudak \& Demorest 2010 who give the profile of the MSP J1012+5307). They have been mentioned in (iv) of Section 1 and the present model gives a natural and direct explanation for their presence in that some approximation to cylindrical symmetry is not unlikely if the observer arc of transit is close to the axis of symmetry. The radiation is also coherent in the sense that the correct retarded phase difference exists between the observer and each point within the current density distribution. The frequency interval to which it refers is, for most pulsars, an order of magnitude smaller than the 400 - 1400 MHz interval of observation. But apart from this, there are some uncertainties in our evaluation.

(i)   Values of $|F|^{2}$ depend strongly on the form of the equal-phase surface of the current density $J$ which is a section of a sphere of radius $2R\eta/3$ to take account of flux-line curvature, and is assumed in equation (9).  

(ii)  For the reasons described in (iv)and (v) of Section 1 the cross-sectional area $\pi u_{m}^{2}$ may be sub-divided into sectors having random relative phases.  In general, this reduces the computed luminosity.

(iii) The current fluctuation in equation (7) may be subject to screening at $u \geq u_{m}$ which has not been considered here.

(iv)  The geometrical form of the system is unusual in that for small observer angle $\theta_{0}$, the electromagnetic radiation passes at radii $u < u_{m}$ through the Langmuir-mode current density prior to leaving the black box at $z_{f}$.  It is a system in which the linear dimension of the current density is large compared with the wavelength radiated. Again, investigation of this, particularly in the case of a non-linear Langmuir mode, lies beyond the scope of the present paper.

A departure from the void black-box is necessary to generate higher frequencies.
The end-point of an unstable Langmuir mode in the electron-positron case has been considered previously by Asseo, Pelletier \& Sol (1990), Asseo (1993) and by Asseo \& Porzio (2006).  These authors also review previous electron-positron work in the field. The ion-proton case is more simple. 
The Langmuir mode in a high-multiplicity electron-positron plasma needs a bimodal velocity distribution whereas in the ion-proton case this exists naturally in $\delta$-function form as a consequence of the charge-to-mass ratio difference. Here we shall assume that growth to non-linearity occurs, with the consequence that equation (7) has to be replaced by a sum of such equations, each for a different Fourier component of the non-linear mode. We shall adopt the most simple assumption, that of a rectangular wave for which the current density fluctuation is replaced by,
\begin{eqnarray}
 \left|\frac{\delta v}{c}\right|\exp(i(k_{0} - \omega t)) \rightarrow \left|\frac{\delta v}{c}\right|\sum_{n} \left(\frac{2i}{n\pi}\right)\exp(in(k_{0} - \omega t)),
\end{eqnarray}
in which $n$ is an odd integer. Computation of the luminosity and spectrum proceed as before giving overlapping bands of emission for adjacent $n$.  The predicted spectral index changes to $-10/3$ for the rectangular wave case, but should lie within the interval $1.3 - 3.3$. 
Circular polarization develops at altitudes $\eta > \eta_{f}$ as described in a previous publication (Jones 2020b).

\section{Conclusions}

The model described in the previous Section provides an elementary physical basis for radio emission that can function only above an ion-proton polar cap in a pulsar with positive polar-cap corotational charge density.  Hence its adoption would fix the nature of the magnetosphere with regard to emissions in the remaining parts of the electromagnetic spectrum. Unfortunately, prediction of luminosities in specific cases does require detailed information about naturally occurring physical systems that is not, and may never be, available.  However, and in a very broad sense, the order of magnitude luminosity obtained in the previous Section is not inconsistent with those listed in the ATNF Catalogue (Manchester et al, 2005), also those observed by Stovall et al (2015) below 100 MHz.

Further investigation of the black box is beyond the scope of this paper but we refer to Asseo (1993) for an extensive discussion of soliton formation and stability in the electron-positron case. One difference which might be relevant here is that the energy of the lowest transverse quantum state is only of the order of 10 eV for a proton as opposed 10 keV for an electron, so that classical proton motion is not obviously impossible.

The present work differs from earlier ion-proton modelling which assumed an unspecified turbulent system (see Jones 2022) emitting radiation isotropically requiring it to have a larger Lorentz factor (10 - 50) for consistency with observed pulse widths. The specific emission process described here is in the frame of the rotating neutron star but at altitudes low enough for aberration to be neglected.

Pulsar observations below 100 MHz are a relatively new addition to the field and it is possible that they will provide significant tests for the ion-proton and electron-positron models.  Flux density measurements have been made by Stovall et al (2015) using 35 to 88 MHz within the Long-Wavelength Array (LWA1 band). These authors describe briefly the difficulties of observation at low frequencies but have measured fluxes for a large number of pulsars.  These are of the order of $10^{2}$ to $10^{3}$ mJy and the broad impression given by Fig. 4 of their paper is that a flux plateau exists at frequencies below 100 MHz which, at higher frequencies, develops into the characteristic decreasing distribution with large negative spectral index. On the low-frequency side there is some evidence for a decline in flux density, usually at about 40 MHz. A possible test of the ion-proton model would be a search for pulsars in which a non-linear  or other development of the Langmuir mode is not reached.  They should be observable below 100 MHz but not at the usual frequencies of 400 and 1400 MHz. 

In considering these peak flux densities at low frequencies, it is difficult to see how the electron-positron mode frequency replacing that of equation (16), which would be larger by some orders of magnitude, could be connected with their origin. This point has been made previously, quite clearly, by Melrose \& Gedalin (1999). The ubiquity of the plasma frequency has to be recognized. It is  closely related with the Langmuir mode frequency.  But even in modelled large-scale bursts of pair creation (see Philippov, Timokhin, \& Spitkovsky 2020) it appears in the Debye length, $c/\omega_{p}$ of the plasma and is related to the frequencies which are predicted.  Dense electron-positron systems have plasma frequencies larger than in the ion-proton case by a factor  $(\kappa m_{p}/m_{e})^{1/2}$ for pair multiplicity $\kappa \approx 10^{3} - 10^{5}$, the usually quoted values.
Consequently, observations below 100 MHz in normal and millisecond pulsars may prove to give productive tests of the emission mechanism.

\section*{Acknowledgments}

It is a pleasure to thank Professor John Miller for several helpful discussions on this paper and the anonymous referee for a very informative review.

\section*{Data availability}

The data underlying this work will be shared on reasonable request to the corresponding author.

\bsp

\label{lastpage}

\end{document}